\documentstyle[12pt]{article}
\begin{document}
\newcommand{\kx}{\kappa}
\newcommand{\sx}{\sigma}
\newcommand{\lx}{\lambda}
\newcommand{\Lx}{\Lambda}
\newcommand{\mt}{\tilde{m}}
\newcommand{\ssa}{\sigma_A}
\newcommand{\ssl}{\sigma_L}
\newcommand{\saa}{\sigma^\alpha_A}
\newcommand{\sal}{\sigma^\alpha_L}
\newcommand{\sbl}{\sigma^\beta_L}
\newcommand{\sslz}{(\sigma_L)_0}
\newcommand{\salz}{(\sigma^\alpha_L)_0}
\newcommand{\sblz}{(\sigma^\beta_L)_0}
\newcommand{\daa}{\delta^{\alpha}_a}
\newcommand{\dbaa}{{\bar{\delta}}^{\alpha}_a}
\newcommand{\DDb}{\bar{\Delta}}
\newcommand{\Vb}{\bar{V}}
\newcommand{\Ub}{\bar{U}}
\newcommand{\Db}{\bar{D}}
\newcommand{\Eb}{\bar{E}}
\newcommand{\Sb}{\bar{\Sigma}}
\newcommand{\Mc}{{\cal{M}}}
\newcommand{\Mct}{\tilde{\cal{M}}}
\newcommand{\beq}{\begin{equation}}
\newcommand{\eeq}{\end{equation}}
\newcommand{\bee}{\begin{eqnarray}}
\newcommand{\eee}{\end{eqnarray}}
\newcommand{\lsim}{\begin{array}{c}<\vspace{-0.32cm}\\\sim\end{array}}
\newcommand{\gsim}{\begin{array}{c}>\vspace{-0.32cm}\\ \sim\end{array}}
\pagestyle{empty}

\begin{flushright}
CERN-TH/95-90\\
OUTP 95-23 P
\end{flushright}
\vspace{3cm}
\begin{center}
{\bf \Large Disoriented and Plastic Soft Terms:\\
 A Dynamical Solution to the Problem \\
of Supersymmetric Flavor Violations }
\\ \vspace{1cm}
S. Dimopoulos,$^{\rm a,}$\footnote{On leave of absence from the
Physics Department, Stanford University, Stanford CA 94305, USA.}
G.F. Giudice$^{\rm a}$\footnote{On leave of absence from INFN, Sezione
di Padova, Padua, Italy.} and
N. Tetradis$^{\rm b}$ \\
\vspace{1cm}
{}$^{\rm a}$Theoretical Physics Division,  CERN\\ CH-1211 Geneva 23,
Switzerland\\[2ex]
{}$^{\rm b}$Theoretical Physics,  University of Oxford\\
1 Keble Road, Oxford OX1 3NP, U.K.\\
\vspace{1cm}

\abstract{
We postulate that the orientation of the soft supersymmetry-breaking terms
in flavor space is not fixed by tree level
physics at the Planck scale; it is a
dynamical variable which depends on fields that
have no tree level potential.
These fields can be thought of as either moduli or as
the Nambu-Goldstone bosons of the spontaneously broken flavor symmetry
which is non-linearly realized by the soft terms. We show that the soft
terms align with the quark and lepton Yukawa couplings,
just as spins align with an
external magnetic field. As a result, the soft terms conserve  individual
lepton numbers and do not cause large flavor or CP violations. The vacuum
adjusts so as to allow  large sparticle splittings to naturally coexist
with flavor conservation. Consequently, the resulting phenomenology is
different from that of minimal supersymmetric theories. We also propose
theories in which the shape of the soft terms in flavor space is
a dynamical variable which depends on fields that
have no tree level potential. This dynamically leads to
partial degeneracy among sparticles and further supression of flavor
violations.
The ideas of this paper suggest a connection
between the space of moduli and the spontaneously broken flavor group.
}
\end{center}
%
\eject


\setlength{\baselineskip}{15pt}
\setlength{\textwidth}{16cm}
\pagestyle{plain}
\setcounter{page}{1}

\newpage

\setcounter{equation}{0}
\renewcommand{\theequation}{{1.}\arabic{equation}}

\section*{1.
Universal versus Disoriented Soft Terms}

The soft supersymmetry(SUSY)-breaking terms \cite{dg,gg}
are important for at least two
reasons.  First,
they are the key ingredient which made the
construction of
realistic supersymmetric theories possible \cite{dg}.  Second, they are
experimentally measurable
quantities since they determine the masses of sparticles.
In early works, motivated by the need to avoid large flavor
violations, it was postulated that soft terms satisfy universality \cite{dg}.
Universality states that the squarks and sleptons
of the three
families are all degenerate in mass at some scale $\sim M_{\rm GUT}$.

Universality has a geometric interpretation which is useful to
appreciate.  To do this,
consider the limit in which all but the gauge couplings of the
supersymmetric standard
model are set to zero.  The resulting theory possesses a $U(3)^5$ global
symmetry which
is called flavor symmetry.  The 3 stands for the number of families and
the  5 for the
number of $SU(3)
\times SU(2) \times U(1)$ superfield
members in a family, which will be labelled by
$A = Q, \bar
U, \bar D, L, \bar E$.  The flavor symmetry is simply a manifestation
of the fact that
gauge forces do not distinguish particles with identical gauge quantum
numbers.
Universality states that the five $3 \times 3$ sparticle squared
mass matrices $m^2_A$
are flavor singlets, {\it i.e.} proportional to the identity.  They are
spheres in flavor
space and  they realize the flavor symmetry in the Wigner mode.
In this paper we wish to suggest an
alternative mechanism to universality
for avoiding large flavor violations.

Let $\Lambda$ be a high energy scale at which supersymmetry breaking
occurs and the soft terms are
determined.  $\Lambda$
can be of the order of the Planck mass $M_{\rm PL}$ -- as in supergravity
-- or smaller, equal to the mass of the messengers that communicate
supersymmetry breaking
to the
ordinary particles.  Our fundamental hypothesis is that physics at the
scale
$\Lambda$ fixes the eigenvalues of the soft terms $m^2_A$
but leaves
their direction in flavor $U(3)^5$ space undetermined.  In other words,
the potential energy $V_{\Lx}$
of the sector which determines the soft
terms at the scale $\Lambda$ is
flavor $U(3)^5$ invariant.  $V_{\Lx}$ does not depend on the $U(3)^5$ angles
which are flat directions of the potential and which will be called here
``moduli". The moduli determine the direction in which the soft terms
point in flavor
space.  They can be thought of as the Goldstone bosons of the flavor
group which is
spontaneously broken by the soft terms $m^2_A$ themselves and
are therefore ``disoriented'' in flavor space.  Therefore,
the simplest way
to state our hypothesis is:  the soft terms realize the flavor symmetry
in the
Goldstone mode.  In contrast, universality states that the soft terms
realize the
flavor symmetry in the Wigner mode.

Our next assumption is that at energies below $\Lambda$ we have the minimal
supersymmetric particle content\footnote{In section 6 we will also discuss
the case of supersymmetric GUTs.} (along with
the decoupled gauge singlet Goldstones/moduli).
We will show that the orientation of
the soft terms is determined by physics at lower energies
-- in particular the flavor-breaking fermion masses -- in a calculable
way.

A simple analogy is to think of the soft terms $m^2_A$ as a spin
$\vec s$ in
space and $U(3)_A$ as ordinary rotational invariance.  The magnitude of
$\vec s$ is
determined by some unspecified ``high energy" dynamics to be non-zero.
This forces rotational invariance to break spontaneously.
$\vec s$  can point in any
direction
until we turn on an external magnetic field $\vec B$
which explicitly breaks
the rotational invariance and forces $\vec s$ to align
parallel to $\vec
B$.  Notice that alignment (or anti-alignment) is preferred and
the maximal subgroup possible, $SO(2)$, is preserved. This completes the
analogy between $\vec s$ and the soft terms on one hand and between
$\vec B$ and the fermion masses on the other.
Perfect alignment would mean that the maximal subgroup consisting of the
product of
all vectorial
$U(1)$ quantum numbers is preserved
and consequently there is no flavor violation.  In the quark
sector since the Kobayashi-Maskawa matrix $K \not= {\bf 1}$ this is not
possible, but the
dynamics will adjust as to reduce flavor violations.

\setcounter{equation}{0}
\renewcommand{\theequation}{{2.}\arabic{equation}}

\section*{2.
Alignment}

Consider the supersymmetric $SU(3) \times SU(2) \times U(1)$  theory with
minimal
particle content, whose gauge interactions possess an $U(3)^5$ global
flavor
symmetry.  As in the standard model, the Yukawa couplings break the
symmetry.  In
addition, flavor symmetry is violated here also by the soft
SUSY-breaking
terms which in general lead to phenomenologically unacceptable
contributions to
flavor-changing neutral current (FCNC) processes.
Let us concentrate first on the soft SUSY-breaking
masses $m^2_A$.
Our hypothesis is that the ${m}^2_A$ are general Hermitian
matrices whose eigenvalues are fixed at the
high scale $\Lx$ where supersymmetry is broken,
but whose
orientation is a dynamical variable determined by
physics below $\Lambda$\footnote{The possibility that
the third generation Yukawa couplings depend on
dynamical variables was considered in ref. \cite{kou}; similar suggestions have
also been proposed in ref.
\cite{nambu}.}.
The soft SUSY-breaking masses ${m}^2_A$ are thus
promoted to fields:
\beq
{m}^2_A \to \Sigma_A \equiv U^\dagger_A \bar \Sigma_A U_A
{}~~~~~~~~~~~A=Q,\bar U ,\bar D ,L, \bar E .
\label{sig}
\eeq
$\bar \Sigma_A$ are diagonal matrices with real, positive
eigenvalues ordered according to increasing magnitude and  $U_A$ are
$3\times 3$ unitary matrices.

Our fundamental hypothesis can now be restated:  $\bar \Sigma_A$ are
fixed by physics at some very high scale $\Lambda$ -- say $\Lambda
\sim M_{\rm PL}$, for concreteness -- whereas $U_A$
are determined only by lower
energy physics,
namely the energetics of the supersymmetric $SU(3) \times SU(2) \times
U(1)$ theory.
For any given $A$, let us write
\beq
U_A
= {\rm exp} \left( i\sum_\alpha \lambda^\alpha\sigma^\alpha_A \right)~,
\label{up}
\eeq
where
$\lambda^\alpha /2$
are the generators of the flavor
group
broken by $\bar \Sigma_A$, in short the six generators of $SU(3)/U(1)^2$.
Thus
$\sigma^\alpha_A$ can be thought of as the Goldstone bosons
of the flavor $U(3)$
group that
has been spontaneously broken by the $\bar \Sigma_A$ VEV.
In reality, the $\sigma^\alpha$ are
pseudo-Goldstone bosons, because quark and lepton masses explicitly
break flavor invariance.
According to our fundamental hypothesis the potential $V_{\Lx}(\saa)$
of the soft terms at the scale $\Lx$
is flat, so that the expectation value of
$\saa$ is undetermined. However, the effective potential at
a lower scales (such as the
supersymmetry scale $m_s$ or the weak scale) receives quantum
corrections from the integration of fluctuations with
characteristic momenta between $\Lx$ and $m_s$.
It is the dynamics of these fluctuations that fixes the value
of $\saa$ in such a way that the soft SUSY-breaking mass
terms are aligned with the Yukawa couplings.

The natural setting in which to carry our discussion is
provided by the approach to
the renormalization group introduced by
Wilson \cite{wilson}.
In his formalism the effects of quantum fluctuations with characteristic
momenta $q^2$ larger than a given cutoff $k^2$ are included
in a $k$-dependent effective action $\Gamma_k$. The scale
$k$ can be viewed as the coarse-graining scale, beyond which
the details of the system are not probed. As a result
fluctuations with characteristic wavelengths smaller than
$2\pi/k$ are integrated out and their effects
are incorporated in the couplings in $\Gamma_k$.
An exact renormalization group equation
describes how the effective action $\Gamma_k$ changes as
the scale $k$ is lowered and the effects of fluctuations
with larger wavelengths are taken into account.

In our problem $k$ can be identified initially
with the high scale $\Lx$ where supersymmetry is broken.
We are interested in the effect of fluctuations
on the shape of the
potential $V_k(\ssa)$ as the scale $k$ is lowered from
$k=\Lx$ to $k=m_s$.
In appendix A we derive the
equation which describes the evolution of
$V_k(\ssa)$. It is
\beq
\frac{\partial V_k(\ssa)}{\partial t} =
- \frac{k^4}{16 \pi^2}
{\rm Str} \log \left[ 1 +\frac{\Mct^2(\ssa,k)}{k^2} \right],
\label{eleven} \eeq
where $t=\log(k/\Lx)$ and $\Mct^2(\ssa,k)$ is the running mass matrix
of the theory. This equation,
when combined with the evolution equation
for the running mass matrix
\beq
\frac{\partial \Mct^2(\ssa,k)}{\partial t} =
\beta_{\Mct} (\xi^i(k)),
\label{twelve} \eeq
describes how the potential evolves as the coarse-graining scale is
lowered and fluctuations with smaller characteristic momenta are
incorporated in it.
The $\beta$-function for the mass matrix can be obtained from the
$\beta$-functions $\beta_{\xi^i}$ of
the running couplings of the theory $\xi^i(k)$.
The boundary conditions
at the scale $k=\Lx$ are given by the assumed (flat)
form of $V_\Lx(\ssa)$ and the tree level form of the
mass matrix $\Mct^2(\ssa,\Lx)=\Mc^2(\ssa)$.
We have not taken into account the fluctuations of the
Goldstone fields $\saa$, despite the fact that they are massless
at tree level.
The reason is that their
contributions to the effective potential
which introduce a non-trivial $\saa$ dependence
are suppressed by powers of
$\Lx$ relative to the ones we have included.
This can be checked in perturbation theory
if we use the fields
${\tilde{\saa}}= \saa \Lx$ which have appropriate mass dimensions
and consider general kinetic and potential terms,
invariant under non-linear realizations of
the $SU(3)/U(1)^2$ symmetry.
The $\beta$-functions $\beta_{\xi^i}$ which are needed for the
calculation of $\beta_{\Mct}$ must be consistently calculated
within the scheme that we have introduced. However,
in an expansion in powers of the couplings
they can be obtained
from standard perturbative calculations \cite{beta}
(at least to one loop, where no scheme dependence is expected).

Let us first consider eq. (\ref{eleven}) with constant
$\Mct^2(\ssa,k)=\Mct^2(\ssa,\Lx)=\Mc^2(\ssa)$, for which it
can be easily integrated.
Keeping the leading terms in $\Lx$ for $k=0$ we find
\beq
V_0(\ssa) = V_\Lx(\ssa) +
\frac{1}{32 \pi^2} \Lx^2 {\rm Str} \Mc^2(\ssa)
+ \frac{1}{64 \pi^2} {\rm Str} \Mc^4(\ssa)
\log \left( \frac{\Mc^2(\ssa)}{\Lx^2} \right).
\label{thirteen} \eeq
This the standard one loop result for the
effective potential.
For most of the evolution from $k=\Lx$ to $k=0$ described by
eq. (\ref{eleven}), we have
$\Mct^2/k^2 \ll 1$ and the logarithm
can be expanded around one, so that
\beq
\frac{\partial V_k(\ssa)}{\partial t} =
- \frac{k^2}{16 \pi^2}
{\rm Str} \Mct^2(\ssa,k)
+ \frac{1}{32 \pi^2}
{\rm Str} \Mct^4(\ssa,k)~~ ...
\label{fourteen} \eeq
For constant $\Mct^2$ this approximation leads to
\beq
V_k(\ssa) = V_\Lx(\ssa) +
\frac{1}{32 \pi^2} (\Lx^2-k^2) {\rm Str} \Mc^2(\ssa)
+ \frac{1}{64 \pi^2} {\rm Str} \Mc^4(\ssa)
\log \left( \frac{k^2}{\Lx^2} \right).
\label{fifteen} \eeq
Comparison with eq. (\ref{thirteen}) indicates that
the part of the evolution with $k^2 \lsim \Mct^2$
simply takes into account threshold effects which
lead to the decoupling of the massive modes.
As a result the masses replace $k^2$ as an effective
infrared cutoff in the logarithm.
The quadratic contribution, on which our discussion is
based, is unaffected by the approximation.
We replace, therefore,
eq. (\ref{eleven}) by eq. (\ref{fourteen})
and neglect the second term in the r.h.s.

We use the perturbative expressions of ref. \cite{beta}
for the $\beta$-functions $\beta_{\xi^i}$.
This is expected to be a good approximation
for the small couplings relevant for our investigation.
The general form of the perturbative $\beta$-function is
\beq
\beta_{\Mct}= \frac{1}{16 \pi^2} \beta_{\Mct}^{(1)}
+ \frac{1}{(16 \pi^2)^2} \beta_{\Mct}^{(2)} ...,
\label{sixteen} \eeq
where only the first two loop contributions are considered.
Starting from the above expression we can iteratively derive an
approximate solution of eq. (\ref{twelve}) for the small
values of $|\Delta t|/16 \pi^2=
|\log(m_s/\Lx)|/16 \pi^2 \simeq (32-37)/16 \pi^2$
which are relevant for our problem.
The leading terms are given by
\bee
\Mct^2(\ssa,k) = \Mc^2(\ssa)
& + \frac{1}{16 \pi^2} \beta_{\Mc}^{(1)} \log \left( \frac{k}{\Lx} \right)
+ \frac{1}{(16 \pi^2)^2} \beta_{\Mc}^{(2)}
\log \left( \frac{k}{\Lx} \right)
\nonumber \\
& + \frac{1 }{2(16 \pi^2)^2}
\left[ \beta_{\xi^i}^{(1)}
\partial \beta_{\Mc}^{(1)}/\partial \xi^i \right]
\log^2 \left( \frac{k}{\Lx} \right) ...
\label{seventeen} \eee
The $\beta$-functions are evaluated at $k=\Lx$ in terms of the
tree-level values of the couplings and masses.
The last term in the second line of the above expression results
from the quadratic term in the Taylor expansion of
$\Mct^2$.
For the quantity in the square brackets summation over $i$ is assumed
and $\beta_{\xi^i}$ (which is in general a matrix) must
be substituted at the point where the derivative with
respect to $\xi^i$ is taken in the expression for
$\beta_{\Mc}^{(1)}$.
The integration of eq. (\ref{fourteen}) is now straightforward.
We find
\beq
V_{m_s}(\ssa) = V_\Lx(\ssa) +
\frac{\Lx^2}{32 \pi^2} {\rm Str} \biggl\{
\Mc^2
- \frac{1}{32 \pi^2}
\beta_{\Mc}^{(1)}
+ \frac{1}{1024 \pi^4}
\left[ \beta_{\xi^i}^{(1)}
\partial \beta_{\Mc}^{(1)}/\partial \xi^i
- 2 \beta_{\Mc}^{(2)} \right]
\biggr\}.
\label{eighteen} \eeq
As we have already pointed out the r.h.s.
of eq. (\ref{eighteen}) must be evaluated in terms of the
tree-level values of the parameters $\xi^i$ at the scale
$\Lx$.
We also need to express the mass matrix of the theory in terms
of $\xi^i$. Our treatment is simplified by the fact
that the Higgs field has not yet developed an expectation
value at $k = \Lx$. The complications arising
from the non-zero Higgs field expectation value
at low scales are neglected in our approximation.
The induced error is small, as the most significant
contributions in the integration of eq. (\ref{fourteen})
come from scales $k \sim \Lx$.

We start by considering the slepton and Higgs mass matrices,
which are given by
\bee
{\Mc}^2_{e,u,d} =
\left(
\matrix{
m^2_{L,Q,Q}  &
0 \cr
{}~ & ~ \cr
0 &
m^2_{\Eb,\Ub,\Db}  \cr}
\right)
\label{nineteen} \eee
and
\bee
{\Mc}^2_{H} =
\left(
\matrix{
m^2_{H_u}+\mu^2  &
B^{\dagger} \cr
{}~ & ~ \cr
B &
m^2_{H_d}+\mu^2  \cr}
\right).
\label{twenty} \eee
The term ${\rm Str} \Mc^2$
in the r.h.s. of eq. (\ref{eighteen})
gives a $\saa$-independent contribution.
The one loop $\beta$-functions for
$m_A$ and $m^2_{H_{u,d}}$ which are relevant for
the discussion of the orientation of the $\saa$ fields
can be obtained from ref. \cite{beta}. We list only the
parts which remain $\saa$-dependent after the trace is
taken:
\bee
{\rm Tr} \beta^{(1)}_{m^2_L} =
&{\rm Tr}  \left[ m^2_L \lx^{\dagger}_e \lx_e
+ 2 \lx^{\dagger}_e m^2_{\Eb} \lx_e
+ \lx^{\dagger}_e \lx_e m^2_L \right] + (\saa-{\rm indep.})
\label{twentyone} \\
{\rm Tr} \beta^{(1)}_{m^2_{\Eb}} =
&{\rm Tr}  \left[ 2 m^2_{\Eb} \lx_e \lx^{\dagger}_e
+ 4 \lx_e m^2_{L} \lx^{\dagger}_e
+ 2 \lx_e \lx^{\dagger}_e m^2_{\Eb} \right] + (\saa-{\rm indep.})
\label{twentytwo} \\
{\rm Tr} \beta^{(1)}_{m^2_{Q}} =
&{\rm Tr}  \left[
m^2_{Q}( \lx^{\dagger}_u \lx_u + \lx^{\dagger}_d \lx_d )
+ ( \lx^{\dagger}_u \lx_u + \lx^{\dagger}_d \lx_d ) m^2_{Q}
+ 2 \lx^{\dagger}_u m^2_{\Ub} \lx_u
+ 2 \lx^{\dagger}_d m^2_{\Db} \lx_d
\right]
\nonumber \\
&~~~~~+ (\saa-{\rm indep.})
\label{twentythree} \\
{\rm Tr} \beta^{(1)}_{m^2_{\Ub}} =
&{\rm Tr}  \left[
2 m^2_{\Ub} \lx_u \lx^{\dagger}_u
+ 4 \lx_u m^2_{Q} \lx^{\dagger}_u
+2 \lx_u \lx^{\dagger}_u m^2_{\Ub} \right]
+ (\saa-{\rm indep.})
\label{twentyfour} \\
{\rm Tr} \beta^{(1)}_{m^2_{\Db}} =
&{\rm Tr}  \left[
2 m^2_{\Db} \lx_d \lx^{\dagger}_d
+ 4 \lx_d m^2_{Q} \lx^{\dagger}_d
+2 \lx_d \lx^{\dagger}_d m^2_{\Db} \right]
+ (\saa-{\rm indep.})
\label{twentyfive} \\
\beta^{(1)}_{m^2_{H_u}} =
&6 {\rm Tr}  \left[
m^2_{Q} \lx^{\dagger}_u \lx_u
+ \lx^{\dagger}_u m^2_{\Ub} \lx_u
 \right]
+ (\saa-{\rm indep.})
\label{twentysix} \\
\beta^{(1)}_{m^2_{H_d}} =
& {\rm Tr}  \left[
6 m^2_{Q} \lx^{\dagger}_d \lx_d
+ 6 \lx^{\dagger}_d m^2_{\Db} \lx_d
+ 2 m^2_{L} \lx^{\dagger}_e \lx_e
+ 2 \lx^{\dagger}_e m^2_{\Eb} \lx_e
 \right]
+ (\saa-{\rm indep.}),
\label{twentyseven} \eee
where $\lx_{e,u,d}$ are the Yukawa matrices.
We choose a basis in which the the matrices
$\lx_{e,u}$ are diagonal and related to the
observable fermion mass matrices at low energies through
$\lx_{e}=m_{e}/v_1$, $\lx_{u}=m_{u}/v_2$, where $v_{1,2}$ are the
the Higgs field expectation values.
Then
$\lx_{d}$
is given by $\lx_{d}=m_{d}K^{\dagger}/v_1$, with $K$
the CKM matrix and $m_d$ diagonal.
The trace of $\beta^{(1)}_\Mc$
can now be easily evaluated, with the result
\bee
{\rm Str} \beta^{(1)}_\Mc = &{\rm Tr}
\left[ 8 \lx^{\dagger}_e \lx_e m^2_L
+ 8 \lx_e \lx^{\dagger}_e m^2_{\Eb}
+ 14 (\lx^{\dagger}_u \lx_u + \lx^{\dagger}_d \lx_d ) m^2_{Q}
+ 14 \lx_u \lx^{\dagger}_u  m^2_{\Ub}
+ 14 \lx_d \lx^{\dagger}_d  m^2_{\Db}
\right].
\nonumber \\
{}~&~
\label{twentyeight} \eee

Let as consider the first term in the r.h.s. of the
above expression.
Starting from the definitions of eqs. (\ref{sig}), (\ref{up})
we find for small $\sigma^\alpha_L$
\beq
{\rm Tr}
(\lx^{\dagger}_e \lx_e
m^2_L)
= -
\sum_\alpha \left( \sigma^\alpha_L \right)^2
\sum_{i>j} | \lx^\alpha_{ij}|^2
\frac{1}{v^2_1}
\left( \Sb_{Li}-\Sb_{Lj} \right)
\left( m_{ei}^2-m_{ej}^2 \right)
+ {\rm {\cal O}} (\sigma^3),
\label{six} \eeq
with $m^2_i$ the charged lepton masses ordered according
to increasing magnitude.
Clearly the effective potential, as it is given by
eqs. (\ref{eighteen}), (\ref{twentyeight}), (\ref{six}) has
a minimum at $\sigma^\alpha_L = 0$.
The same conclusion can be easily reached for
$A=\Eb$.
The result $U_L = U_{\bar E} = {\bf 1}$ has the important consequence
that the $e, \mu,
\tau$ lepton numbers are separately conserved.  Since slepton and lepton
mass
matrices are parallel, they both preserve the same $U(1)^3$ symmetry and
individual
lepton number violating processes like $\mu \rightarrow e \gamma$ do not
occur in
this theory.
Complete alignment of the squark-quark
sectors is not possible due to the presence of
the Kobayashi-Maskawa matrix.
The soft terms $m^2_{\Ub}$, $m^2_{\Db}$
align with the quark mass matrices $m^2_u$, $m^2_d$ respectively,
while $m^2_Q$ aligns with the linear combination
$m^2_u + K m^2_d K^{\dagger}$.
In this way FCNC processes are adequately suppressed.

The predicted masses of the pseudo-Goldstone bosons
$\saa$ can be computed
from the effective potential of
eqs. (\ref{eighteen}), (\ref{twentyeight}),
(\ref{six}). We find the approximate expression
\beq
m^2_{\saa} = \frac{m^2_s}{(16 \pi^2)^2}
\sum_{i>j} | \lx^\alpha_{ij}|^2
\frac{(m_{Ai}^2-m_{Aj}^2)}{m^2_{\rm weak}},
\label{twentynine} \eeq
where $m_{Ai}$ are the masses of fermions
in the superfield $A$, and
we have assumed that the sparticle splittings are
of the order $m_s$ and approximated the
Kobayashi-Maskawa matrix by
the unit matrix.
The pseudo-Goldstone
masses are proportional to the scale of the soft terms
$m_s$, and to the Yukawa couplings,
which explicitly break the flavor symmetry.
They range roughly between $(10^{-3}-1)$ GeV.
The couplings of the pseudo-Goldstone bosons to the
particles of the minimal supersymmetric standard model
are extremely weak as they are suppressed by $m_s/\Lx$.

The quadratic momentum dependence in
eq. (\ref{fourteen}) shows that the alignment is determined
by the behavior of the theory at energies just below the scale $\Lx$.
This feature is not appealing since it introduces a sensitivity
to the details of the high-energy physics. In view of this it is not
appropriate to think of the Goldstones/moduli  as determined by low-energy
physics. Unfortunately this ultraviolet sensitivity  is bound to
frustrate all attempts to convert parameters of the supersymmetric
theory -- through their dependence on moduli -- to dynamical variables
of the low-energy theory. It originates in the quadratic dependence of
the energy on the cut-off, a feature present in theories of softly-broken
low-energy supersymmetry\footnote{Similar observations were made in ref.
\cite{bag}.}.

\setcounter{equation}{0}
\renewcommand{\theequation}{{3.}\arabic{equation}}

\section*{3.
Alignment of the $A$-terms}

The triscalar $A$-terms
break both supersymmetry and chirality; thus they resemble the
soft masses $m^2_A$ in one sense and the Yukawa couplings in another.
Consequently there are three possibilities: \\
a) The first is that
the $A$-terms are disoriented and independent of
$m^2_A$. They can be parametrized as
\beq
A_a = V_a \DDb_a \Vb_a,
\label{threeone} \eeq
with $a = e,u,d$. $\DDb_a$ is the diagonal matrix of the
eigenvalues of $A_a$, and $V_a$, $\Vb_a$ are unitary
matrices, analogous to $U_A$, which include new Goldstone
fields $\delta^{\alpha}_a$, ${\bar{\delta}}^{\alpha}_a$
whose values are postulated to be undetermined
by the potential $V_\Lx$ at the high scale $\Lx$.
Their expectation values at low scales are determined by the minimum of
the potential of eq. (\ref{eighteen}), in which the additional
fields $\daa$, $\dbaa$ now appear.
The term proportional to ${\rm Str} \beta_{\Mc}^{(1)}$
in the r.h.s. of eq. (\ref{eighteen}) is $A_a$-independent
and does not determine $\daa$, $\dbaa$. One has to evaluate
the last two terms
which do depend on $A_a$ and can fix the values of
$\daa$, $\dbaa$. Making use of the results of
ref. \cite{beta} for $\beta_{\xi^i}^{(1)}$, $\beta_{\xi^i}^{(2)}$
we conclude that all the terms which depend quadratically
on $A_a$, e.g. ${\rm Tr} A^{\dagger} A \lx^{\dagger} \lx$,
come with a positive sign in the potential
and anti-align the $A$-terms with the
Yukawa couplings. This still implies the existence of a $U(1)^3$
symmetry -- approximate for quarks, exact for leptons --
which suppresses flavor violations. However, the terms linear
in $A_a$, e.g. ${\rm Tr} A \lx^{\dagger}$, which are
proportional to the gaugino masses, align the $A$-terms with
the Yukawa couplings, irrespectively of their sign.
Therefore, depending on whether the terms linear
or quadratic in $A_a$ dominate we expect alignment or
anti-alignment. Either possibility guarantees flavor conservation
as either one implies an approximate $U(1)^3$ symmetry.
The former situation occurs if the gaugino mass is much bigger
than the $A$-terms; and the latter in the opposite case. \\
b) A second possibility is that the orientation of the $A$-terms
is given by the same matrices $U_A$ that occur in $m^2_A$, i.e.
\beq
A_e = U^{\dagger}_L \DDb_e U_{\Eb}~~~~~~~~~~
A_u = U^{\dagger}_Q \DDb_u U_{\Ub}~~~~~~~~~~
A_d = U^{\dagger}_Q \DDb_d U_{\Db}.
\label{threetwo} \eeq
In this case the orientation
of the $U_A$ is fixed by the dominant one loop effects of the previous
section and one has to hope that the frozen parameters $\DDb_a$
commute with the corresponding mass matrices $m_a$.\\
c) Finally, a third possibility is that the $A_a$ themselves
are frozen parameters; this is identical to what happens in the
minimal supersymmetric standard model and one has again to hope
that the $A_a$ commute with the corresponding $\lx_a$.

For the rest of the paper we shall assume that the $A$-terms
commute with the corresponding Yukawa matrices and do not
cause significant flavor violations.

\setcounter{equation}{0}
\renewcommand{\theequation}{{4.}\arabic{equation}}

\section*{4.
Flavor Violating Processes}

The first consequence of the results of the previous sections is
that, as a result of alignment, all three lepton
numbers are
individually conserved.
This is obviously not possible in the quark sector, since the up and
down quarks
themselves do not have parallel mass matrices\footnote{Alignment as a solution
of the flavor problem in supersymmetic theories was also considered,
in a different context, in ref. \cite{nir}.}.  The quark flavor
violations are best
discussed by going, via a superfield rotation, to the quark Yukawa
eigenbasis where both
up and down masses are diagonal and the squark Yukawas have the form:
\bee
{\cal M}^2_u =&
\pmatrix{
m^2_u + S^\dagger \bar \Sigma_Q S + D_{u_L} & \bar \Delta_u +
\frac{\mu}{\tan\beta} m_u \cr \cr
\bar \Delta_u + \frac{\mu}{\tan\beta} m_u & m_u^2  +
\bar \Sigma_{\bar U} +
D_{u_R}}
\nonumber \\
{\cal M}^2_d =&
\pmatrix{
m^2_d + K^\dagger S^\dagger \bar \Sigma_Q SK + D_{d_L}&
\bar \Delta_d + \mu\tan\beta
m_d\cr \cr
\bar \Delta_d + \mu \tan\beta m_d & m^2_d + \bar \Sigma_{\bar D} + D_{d_R}}
\label{fmas}
\eee
All flavor violation is contained in $S$ and $K$.  The off-diagonal
elements of $S$
are much smaller than those of the Kobayashi-Maskawa matrix:
\bee
S_{23} &\simeq& K_{cb} \frac{m^2_b}{m^2_t} \sim 2 \times 10^{-5}
\nonumber \\
S_{13} &\simeq& K_{ub} \frac{m^2_b}{m^2_t} \sim 2 \times 10^{-6}
\nonumber \\
S_{12} &\simeq& \frac{|K_{us}K^\star_{cs} m^2_s + K_{ub}K^\star_{cb}
m^2_b|}{m^2_c} \sim 5
\times 10^{-3}~,
\eee
and therefore they do not significantly affect FCNC processes, although
they may contribute to CP-violating processes.
Then, in the approximation $S = {\bf 1}$, all new flavor
violations occur
in the $D_L$ sector, as can be seen from the squark mass matrices in Eq.
(\ref{fmas}).

The most stringent constraint comes from the contribution of squark-gluino
loops to the real
part of the $K^0-\bar K^0$ mixing:
\beq
\left( \frac{\Delta m_K}{m_K} \right)_{\tilde g} =
\frac{f^2_KB_K}{54}~\frac{\alpha^2_s}{M_{\tilde{g}}^2}~{\rm Re} (X)
\label{kk}
\eeq
\beq
X\equiv \sum_{i,j} K_{is}K_{id}^\star K_{js}K_{jd}^\star ~
f\left(\frac{{m}^2_{Q_i}}
{M^2_{\tilde g}}, \frac{{m}^2_{Q_j}}
{M^2_{\tilde g}}\right)
\eeq
\beq
f(x,y)\equiv \frac{1}{x-y}\left[ \frac{(11x+4)x}{(x-1)^2}\log x -\frac{15}{x-1}
- (x\to y)\right] ,
\eeq
where $f_K=165$ MeV is the kaon decay constant, $B_K$ parametrizes the hadronic
matrix element, and $M_{\tilde g}$ is the gluino mass.
Assuming $M^2_{\tilde g} = {m}^2_Q$ and keeping the
leading contribution in the squark mass splitting, one finds
\beq
{\rm Re} (X) = \frac{\sin^2 \theta_c}{6} {\cal D}_{21}^2~,
\eeq
where $\theta_c$ is the Cabibbo angle and
\beq
{\cal D}_{ij} \equiv
\frac{ m^2_{Q_i} -  m^2_{Q_j}}{
m^2_{Q_i}}~.
\eeq
If we require that the gluino contribution in Eq. (\ref{kk}) does not exceed
the experimental value of $\Delta m_K/m_K$, we obtain the constraint:
\beq
{\cal D}_{21} < 0.1~\frac{ m_Q}{300~\hbox{GeV}}~.
\label{kkkk}
\eeq

The squark-gluino contribution to the imaginary part of $K^0-\bar K^0$
mixing is given by:
\beq
(|\epsilon|)_{\tilde g} =
\frac{m_K}{\Delta m_K}~ \frac{f^2_KB_K}{108 \sqrt 2} ~
\frac{\alpha_s^2}{M^2_{\tilde{g}}}
{}~{\rm Im}(X)
\eeq
With the same approximation used before, we obtain
\beq
{\rm Im} (X) = \frac{1}{3}
|K_{us}|
|K_{ub}|
|K_{cb}| \sin \delta ~{\cal D}_{32} {\cal D}_{21}~,
\eeq
where $\delta$ is the CP-violating phase in the Kobayashi--Maskawa matrix.
This does not exceed the experimental value for $|\epsilon|$ if
\beq
\sqrt{ {\cal D}_{21} {\cal D}_{31} } <  \frac{ m_Q}{300~ \hbox{GeV}}~.
\eeq

There is no significant constraint coming from $B^0-\bar B^0$
mixing and, in the limit $S={\bf 1}$, there is no new gluino-mediated
contribution to $D^0-\bar D^0$ mixing.

The constraints from FCNC on our model are much milder than those on a
general supersymmetric $SU(3)
\times SU(2) \times U(1)$ theory with minimal particle content and
non-universal frozen soft-terms \cite{fcnc}.
The reason is that in our theory, just as in the standard model,
flavor violations
are proportional to the Kobayashi-Maskawa
angles;  however, they are also suppressed by the large sparticle
masses.  Therefore, our
contributions to rare processes can compete with the standard model
contributions
only if the latter have light
quark suppressions, as in
$\Delta m_K/m_K$ where $(\Delta m_K/m_K)_{SM} \sim G_Fm^2_c$.

It is noteworthy that we do not obtain any constraints from either $\mu
\rightarrow e\gamma$ or
$\epsilon$.  These provide by far the strongest constraints on general
supersymmetric models.  In
our case, $\mu \rightarrow e\gamma$ vanishes whereas $\epsilon$ is
small because it is
proportional to the Jarlskog invariant $J$ of the standard model
and is further
suppressed by sparticle masses.
The only significant constraint we have is from $\Delta m_K$ Eq.~(\ref{kkkk}).
  It can be
accounted for in several ways.
One is by invoking heavy gluinos, which cause the squark masses to
approach  one another in the
infrared.  Furthermore, in Sect. 5, we will show how the dynamics of
the moduli can adjust to
render the squarks of the two heavy generations degenerate.

We end with a cosmological caveat.
Because the
moduli couple
very weakly with strength $ \sim M^{-1}_{\rm PL}$, they do not
efficiently lose energy. As a result, they do not reach
their minima in simple cosmologies \cite{modpr}, unless they
happen
to accidentally start out near their vacuum.
Recently, there have been a revival of suggestions
\cite{modsol} on how to solve the problem and to allow the
moduli to
cosmologically relax to their ground state.  Such a mechanism is clearly
necessary to
ensure flavor alignment. Even more, it
is necessary
to ensure that the Universe is not overclosed by coherent oscillations
of the moduli.

\setcounter{equation}{0}
\renewcommand{\theequation}{{5.}\arabic{equation}}

\section*{5.
Plastic Soft Terms}

In previous sections we have conjectured that the potential
$V_{\Lambda}$
at the scale $\Lx$ where SUSY is broken
leaves the orientation of the
soft terms undetermined, but fixes their eigenvalues.
In this section we wish to relax the latter hypothesis.
We envisage that the supersymmetry-breaking dynamics at $\Lambda$ provide the
low-energy theory with a
constraint which fixes the overall scale $m_s$ but does not necessarily
freeze all three
eigenvalues.  Some functions of the eigenvalues can correspond to
flat directions which remain
undetermined until we turn on the Yukawa couplings. Of course, our
postulate that the
supersymmetry-breaking
mechanism respects the flavor symmetry requires that the
constraints that fix $m_s$
have to be flavor singlets.

Let us consider the case of vanishing left-right mixings in the
squark and
slepton mass matrices and focus on the fields $\Sigma$ defined
in eq. (\ref{sig}). (For this section we drop the subscript
$A$.)
Suppose that the dynamics at the scale $\Lambda$ fixes the two
lowest-dimension flavor-singlet operators:
\beq
{\rm Tr}\Sigma = T~, \quad {\rm Tr}\Sigma^2 = T_2~,
\label{trace}
\eeq
where $T^2$ and $T_2$ are numbers of order $m^4_s$.

These are two constraints on three eigenvalues, thus one combination of
eigenvalues remains a flat
direction whose VEV will be determined by low-energy physics in a
calculable way.  It is easy to
identify the flat direction.  The above constraints are not just $SU(3)$
invariant, but are
$SO(8)$ invariant, and they force the spontaneous breakdown
$SO(8) \rightarrow SO(7)$, giving
rise to seven Goldstone bosons.  Six of them are a consequence of the
breaking $SU(3) \rightarrow
U(1)^2$ and can be identified with the fields $\sigma$.  The
seventh is the new flat direction $\theta$ which allows the
eigenvalues of $\Sigma$ to slide
along a valley which preserves the above constraints.

The field $\Sigma$ satisfying Eq. (\ref{trace}) can be expressed as
\beq
\Sigma =  T U^\dagger  \left[ \frac{1}{3} - x(\cos \theta \lambda_8 + \sin
\theta \lambda_3)
\right]U~,
\eeq
where $\lambda_{3,8}$ are the two diagonal Gell-Mann matrices, $U$
denotes an
$SU(3)/U(1)^2$ rotation, and
\beq
x \equiv \sqrt{\frac{3T_2 - T^2}{6T^2}}~, \quad 0 \leq x \leq
\frac{1}{\sqrt3}~.
\eeq
Our assumption is that the six parameters contained in $U$ and the angle
$\theta$ are
dynamical variables, related to flat directions of the moduli fields.
The soft term $\Sigma$ is not only ``disoriented" in flavor space,
but is also ``plastic", since the pattern of eigenvalues can be deformed.
Plasticity is disorientation in $SO(8)$ space. In contrast to
$SU(3)$, $SO(8)$ allows rotations in the $\lx_3 - \lx_8$ plane.

The effective
potential for $\Sigma$
is given by eqs. (\ref{eighteen}), (\ref{twentyeight}), and
its minimum occurs for $\cos \theta \simeq 1.$
This implies that the vacuum has an approximate $SU(2) \times U(1)$ symmetry
which insures the degeneracy of the soft masses
of same charge sparticles belonging to the two lightest generations.
Consequently, $m^2_{Q_1}$ and $m^2_{Q_2}$ are approximately equal, and this
provides for the desired suppression of the real part
of $K_0 - \bar{K}_0$ mixing.
Plasticity can be extended to the $A$-terms and, as
discussed in section 3, they can align or anti-align
with the Yukawas,
depending on whether the gaugino mass is much larger
than the $A$-terms or vice versa.

\setcounter{equation}{0}
\renewcommand{\theequation}{{6.}\arabic{equation}}

\section*{6.
Minimal Unification}

Until now we have been working under the hypothesis that below the scale
$\Lambda$, where the supersymmetry breakdown occurs, we have the minimal
supersymmetric $SU(3)\times SU(2)\times U(1)$ particle content. We now
consider the possibility that the theory below $\Lambda$ is some minimal
supersymmetric GUT.

In minimal supersymmetric GUTs the gauge symmetry is increased to $SU(5)$ or
$SO(10)$ and the
number of chiral multiplets decreases.  This means that the flavor
group is no
longer $U(3)^5$, but it is smaller: $U(3)_{\bf    {\bar 5}} \times
U(3)_{\bf  10}$ in the case of $SU(5)$, and just $U(3)_{\bf 16}$
for $SO(10)$.
If we also
assume that the soft terms are as minimal as possible, namely singlets
under the GUT
group, then we have a very constrained system with a small flavor group
and a small
number of parameters in the soft terms.  Are there enough
moduli/Goldstones
available to align sufficiently and avoid problems with flavor
violations ?

For
simplicity, let us discuss the minimal $SO(10)$ model in which the
Yukawa coupling superpotential between the ordinary fermions in the
${\bf 16}$ representation and the Higgs fields $H_{u,d}$ is
\beq
W_Y = {\bf 16} \lambda_u {\bf 16} H_u + {\bf 16} K \lambda_d K^T {\bf 16}
 H_d~.
\eeq
For simplicity, we will ignore the $A$ trilinear terms and write the
soft supersymmetry-breaking Lagrangian as:
\beq
{\cal L}_{\rm Soft} = \frac{m^2_s}{F} {\bf 16}^\dagger
U^\dagger \bar \Sigma U{\bf 16}~.
\label{gut}
\eeq

The crucial difference between this minimal-GUT case, with gauge-singlet
$\Sigma$, and the previous $SU(3)\times SU(2)\times U(1)$ analysis is
apparent from Eq. (\ref{gut}). Now there is just one $U$ available,
instead of 5, to do all the alignments necessary to reduce flavor
violations. It is clear that $U$
will align $\Sigma$ parallel to $m_u$, since $m_u$ gives the largest
contribution to the energy.
This implies that all
sparticle mass matrices will be parallel to $m_u$, whereas the down-quark
and charged-lepton mass matrices will be misaligned from $m_u$ by angles
of the order of the Kobayashi-Maskawa angles.

Thus, unless sleptons are highly degenerate in mass,
$\mu_{L,R} \rightarrow e_{R,L}\gamma$ transitions are
proportional to a
mixing angle $K_{e\mu} = K_{us} \simeq \sin \theta_c$ and occur at an
unacceptable rate.  In $SU(5)$  only
the
right-handed sleptons are misaligned from the lepton mass matrix, and
the amplitude for
$\mu_L \rightarrow e_R + \gamma$ is again proportional to the Cabibbo
angle $\sin \theta_c
\simeq
\sqrt{d/s}$.
Of course, minimal $SO(10)$ and $SU(5)$ theories have a problem:  they
predict $m_d =
m_e$ and this is the reason why they give $\mu \rightarrow e\gamma$
proportional to
$\sqrt{d/s}$. However even if we extend the theory {\it \`a la}
Georgi--Jarlskog, the $\mu \rightarrow e\gamma$ amplitude is still
problematic, being
proportional to
$\sqrt{e/\mu}$.

The reason for this failure is that in minimal supersymmetric GUTs with
minimal
GUT-invariant soft terms, the few available soft terms just align with
$m_u$, leaving some mismatch between down quarks and squarks
and more importantly between leptons and sleptons.
This causes difficulties with individual lepton violating processes, which
were not originally present in supersymmetric GUTs with universality at
$M_{GUT}$.

The problem could be cured in more complicated GUTs with a larger flavor
structure,  necessary perhaps to explain the fermion mass pattern, which
would allow for more freedom in the low-energy alignment of the soft-breaking
masses.

A strong degeneracy between the first two generations of sleptons and
down squarks suppresses the most dangerous processes and could therefore
represent an alternative solution. In the previous section we have
shown that this occurs in the plastic soft-term scenario
and the degeneracy of the sparticles of the first
two generations is {\it predicted}. The dynamics of the plastic soft terms
cures the disease in the
dynamics of the disoriented soft terms: in GUTs the large up-type quark Yukawa
couplings force the
sleptons to misalign, but insure that the first two generations are
almost degenerate in mass.
The decay $\mu \to e \gamma$ can still occur via virtual $\tilde{\tau}$
exchange, and its rate is just below the present experimental limit.
Interesting effects in lepton-number violating $\tau$-decays can be
present. The plastic GUT scenario allows therefore the construction of
phenomenologically viable models which are predictive and represent
possible alternatives to the minimal supersymmetric standard model with
universal boundary conditions at $M_{GUT}$.

\setcounter{equation}{0}
\renewcommand{\theequation}{{7.}\arabic{equation}}

\section*{7.
Conclusions}

We proposed ``disorientation" as an alternative to universality for
suppressing flavor violation in supersymmetric theories. Universal
soft terms realize the flavor symmetry in the Wigner mode. Disoriented
soft terms realize it in the Nambu-Goldstone mode; this allows large
sparticle splittings and has the appeal that the absence of flavor
violations is a consequence of a dynamical calculation.

The Goldstone particles can be thought of as either the consequence
of a spontaneously broken flavor symmetry or perhaps could be
identified with some of the flat directions (moduli) that frequently
occur in supersymmetric or superstring theories. In the latter case
there would be an important connection between the space of the moduli
and the flavor group.

Why did our mechanism work? Promoting some of the parameters of the
low-energy theory to fields allowed us to exploit nature's preference
for states of maximal possible symmetry.
This is the reason why:
 the spin aligns with an external magnetic field, preserving SO(2);
  sleptons align with leptons, preserving individual lepton number
conservation  $U(1)^3$;
 squarks align  --as much as possible-- with the quarks, preserving
an approximate $U(1)^3$;
   the 7th goldstone boson of the plastic scenario  chooses to relax
at its special value where the symmetry is enhanced to $SU(2) \times
U(1)$ and pairs of sparticles are degenerate.
Nature's frequent preference for states of higher symmetry fully
accounts for our mechanism for the suppression of flavor violation.
More importantly, it leads us to
new supersymmetric phenomenology and the peaceful coexistence of
split sparticles and flavor conservation.

We have found that the dynamics of alignment occurs at
large scales and is sensitive to details of Planckian physics. In
light of this, is disorientation better than universality?
Both are strong hypotheses
which rely on the existence of an approximate symmetry
in a sub-sector of the full theory.
Which is better can only be decided in the context
of a complete theory
which addresses the full flavor problem and explains
fermion masses.
Only then can we see how the soft terms avoid being directly
infested with large flavor violations from Planckian physics.

\vspace{0.3cm}
\noindent
{\bf Acknowledgements:}

It is a pleasure to thank L. Hall, C. Kounnas, G. Veneziano and
F. Zwirner for valuable discussions.
S.D. thanks the
Department of Theoretical Physics of the
University of Oxford for hospitality
during the course of the work.

\newpage

\setcounter{equation}{0}
\renewcommand{\theequation}{{A.}\arabic{equation}}

\section*{Appendix A: The evolution equation for $V_k$}

We are interested in the effect of fluctuations
on the shape of the
potential $V_k(\ssa)$ as the scale $k$ is lowered from
$k=\Lx$ to $k=m_s$.
For this reason we
introduce an effective infrared cutoff term $R_k(q^2)$
in the momentum integrations appearing in the loop
contributions to the potential
$V_k(\ssa)$. This term prevents
the integration of modes with momenta $q^2 \lsim k^2$.
The effective potential at one loop
is now given by
\beq
V_k(\ssa) = V_\Lx(\ssa)
+ \frac{1}{2} \int_\Lx \frac{d^4q}{(2 \pi)^4}
{\rm Str} \log \left[ q^2 + R_k(q^2) + \Mc^2(\ssa) \right].
\label{seven} \eeq
In the formulation by C. Wetterich
\cite{exact} the cutoff term is chosen as
\beq
R_{k}(q^2) = \frac{Z_{k} q^2 f^2_k(q^2)}{1- f^2_k(q^2)}.
\label{eight}
\eeq
The function
\beq
f^2_k(x) = \exp \biggl\{
-2 a \left( \frac{q^2}{k^2} \right)^b \biggr\}
\label{nine}
\eeq
can be used for the
implementation of a sharp or smooth cutoff through
an appropriate choice of the two free parameters $a,b$.
$Z_k$ is a $k$-dependent matrix in field space whose
precise definition is given in the following.
An ultraviolet cutoff $\Lx$ is assumed for the momentum integration.
Notice that for $k=\Lx$ the one loop contribution automatically
vanishes. In the limit $k \rightarrow 0$ the
cutoff term $R_k(q^2)$ is removed and the integration
reproduces the standard one loop result for the effective potential
without a cutoff.
Taking the partial derivative of $V_k$ with respect to
$t=\log(k/\Lx)$ results in the evolution equation
\beq
\frac{\partial V_k(\ssa)}{\partial t} =
\frac{1}{2} \int \frac{d^4q}{(2 \pi)^4} {\rm Str}
\frac{\partial R_k}{\partial t}
\left[ q^2 + R_{k}(q^2) + \Mc^2(\ssa) \right]^{-1}.
\label{ten} \eeq
The momentum integration is infrared and ultraviolet finite as the
integrand deviates significantly from zero only for $q^2 \simeq k^2$.
The renormalization group improvement consists in substituting
the running mass matrix $\Mc^2(\ssa,k)$ for the classical one,
and multiplying $q^2$ by the wavefunction
renormalization of the various fields $Z_{k}$.
This takes into account the fact that the change in the
effective potential when fluctuations with momenta $q^2 \simeq k^2$
are incorporated in it
involves the full propagator of the theory
at the scale $k$.
We can now identify the matrix $Z_k$ appearing in the
definition of eq. (\ref{eight}) with the wavefunction renormalization.
Notice that the $t$-derivative of $R_{k}$
includes a contribution proportional to the anomalous dimension
of the fields $\eta=-\partial (\log Z_{k})/ \partial t$.
It can be checked that the explicit
$Z_k$-dependence can be incorporated in the definition of
the renormalized mass matrix
$\Mct^2(\ssa,k) = Z_k^{-1} \Mc^2(\ssa,k)$.
The integral in the r.h.s. of eq. (\ref{ten}) cannot
be easily computed for general values of the parameters
$a,b$ appearing in eq. (\ref{nine}). However, in the limit
of a sharp cutoff $b \rightarrow \infty$ the momentum integration
can be carried out explicitly. Moreover, the contribution
proportional to $\eta$ can be neglected, as it is suppressed by $1/b$.
As a result, the effect of the wavefunction renormalization
is completely absorbed in the running of the renormalized mass
matrix $\Mct(\ssa,k)$.
The final expression for the running of the potential is
\beq
\frac{\partial V_k(\ssa)}{\partial t} =
- \frac{k^4}{16 \pi^2}
{\rm Str} \log \left[ 1 +\frac{\Mct^2(\ssa,k)}{k^2} \right].
\label{appe} \eeq

A few remarks are due in order to clarify
some steps in our derivation: \\
1) Even though
we were led to eq. (\ref{appe}) through an intuitive way
a more formal derivation is possible \cite{exact}. The
$k$-dependent effective action $\Gamma_k$ for scalar fields
can be obtained from the partition function
through the usual Legendre transformation, if the infrared cutoff
term of eq. (\ref{eight})
is added to the classical action so that low momentum modes
do not propagate.
An exact renormalization group equation describes the
evolution of $\Gamma_k$ with $k$ \cite{exact}. This equation
leads to eq. (\ref{appe}) for the potential.
For fermions the discussion proceeds along parallel lines.
A modified fermion propagator is used, so that the momentum
integrations are cut off in the infrared \cite{fermions}.
For the discussion of supersymmetric theories the choice
of cutoffs terms for scalars and fermions must preserve the
supersymmetry at all scales. This is accomplished
in the limit $b \rightarrow \infty$ that we have
considered \cite{fermions}. \\
2) The matrix $Z_k$ appearing in eq. (\ref{eight})
includes the wavefunction renormalization for scalar and
fermion fields. We have implicitely assumed
that the fermionic part of the matrix involves the square of
the term which renormalizes the fermion field. This is apparent
from the way the renormalized masses $\Mct$ are defined. \\

\end{document}